\documentclass[11pt]{article}
\usepackage{amssymb,latexsym,amsmath,amsbsy}
\usepackage[dvips]{graphicx}
\usepackage{cite}
\newtheorem{theo}{Theorem}
\newtheorem{defi}[theo]{Definition}

\def\ds{\displaystyle}
\def\nn{\nonumber}

\def\ch{\mathop{\rm char}\nolimits}

\def\wt{\mathop{\rm wt}\nolimits}
\def\qdots{\mathinner{\mkern1mu\raise1pt\vbox{\kern7pt\hbox{.}}\mkern2mu
 \raise4pt\hbox{.}\mkern2mu\raise7pt\hbox{.}\mkern1mu}}
\def\Z{{\mathbb Z}}

\def\C{{\mathbb C}}
\def\gl{\mathfrak{gl}}

\def\u{\mathfrak{u}}
\def\g{\mathfrak{g}}
\def\h{\mathfrak{h}}
\def\n{\mathfrak{n}}
\def\p{\mathfrak{p}}
\def\so{\mathfrak{so}}

\def\osp{\mathfrak{osp}}
\newcommand{\s}[1]{\langle #1 \rangle}

\def\mybox{\hfill$\Box$}
\begin{document}
\begin{center}
{\Large \bf
Parafermions, parabosons and representations\\[2mm]
 of $\so(\infty)$ and $\osp(1|\infty)$ }\\[5mm]
{\bf N.I.~Stoilova}\footnote{E-mail: Neli.Stoilova@UGent.be; Permanent address:
Institute for Nuclear Research and Nuclear Energy, Boul.\ Tsarigradsko Chaussee 72,
1784 Sofia, Bulgaria} {\bf and J.\ Van der Jeugt}\footnote{E-mail:
Joris.VanderJeugt@UGent.be}\\[1mm]
Department of Applied Mathematics and Computer Science,
Ghent University,\\
Krijgslaan 281-S9, B-9000 Gent, Belgium.
\end{center}


\begin{abstract}
The goal of this paper is to give an explicit construction of the Fock spaces
of the parafermion and the paraboson algebra, for an infinite set of generators.
This is equivalent to constructing certain unitary irreducible lowest weight representations
of the (infinite rank) Lie algebra $\so(\infty)$ and of the Lie superalgebra $\osp(1|\infty)$.
A complete solution to the problem is presented, in which the Fock spaces have basis vectors
labelled by certain infinite but stable Gelfand-Zetlin patterns, and the transformation
of the basis is given explicitly.
We also present expressions for the character of the Fock space representations.
\end{abstract}

\setcounter{equation}{0}
\section{Introduction} \label{sec:Introduction}%

Parafermions and parabosons~\cite{Green} arose in an attempt to generalize the second quantization
method in the case of the free field, by permitting more general kind of statistics than the
common Fermi-Dirac and Bose-Einstein statistics.
These generalizations have an algebraic formulation in terms of generators 
and relations~\cite{Green,Greenberg,Ohnuki,Volkov}.
The parafermion operators $f_j^\pm$, $j=1,2,\ldots$ (generalizing the
statistics of spinor fields), satisfy certain triple
commutation relations, see~\eqref{f-rels}, and it has been known for a long time~\cite{Kamefuchi,Ryan}
that the Lie algebra generated by these operators subject to the defining triple relations
is an orthogonal Lie algebra. More precisely, for a finite set of parafermions ($j=1,2,\ldots,n$)
it is the Lie algebra $\so(2n+1)$. 
In the context of quantum field theory, however, one is mainly interested in the
case of an infinite set of parafermions, $j\in\Z_+=\{1,2,\ldots\}$.
The corresponding parafermion algebra is an infinite rank Lie algebra,
often referred to as $\so(\infty)$.

The paraboson operators $b_j^\pm$, $j=1,2,\ldots$ (generalizing the
statistics of tensor fields), also satisfy certain triple
relations, see~\eqref{b-rels}, this time however involving anticommutators and commutators.
It was therefore logical to look for a connection with Lie superalgebras instead of 
with Lie algebras. 
This connection was established in~\cite{Ganchev}, and the Lie superalgebra generated by 
a finite set ($j=1,2,\ldots,n$) of paraboson operators subject to the defining triple relations
is the orthosymplectic Lie superalgebra $\osp(1|2n)$. 
For an infinite set of parabosons, $j\in\Z_+=\{1,2,\ldots\}$,
the paraboson algebra is an infinite rank Lie superalgebra denoted as $\osp(1|\infty)$.

The main objects of interest are the unitary lowest weight representations of the
relevant algebras with a nondegenerate lowest weight space (i.e.\ with a unique
vacuum). 
These representations are precisely the so-called Fock spaces of the parafermion
or paraboson algebras.
These Fock spaces are characterized by a positive integer~$p$, often referred to 
as the order of statistics.
The Fock spaces $W(p)$ (for parafermions) and $V(p)$ (for parabosons) can be defined
in many ways, and one of the standard methods is due to Green and
known as Green's ansatz~\cite{Green,Greenberg}. 
The computational difficulties of Green's ansatz are
related to finding a proper basis of an irreducible constituent of a $p$-fold
tensor product~\cite{Greenberg}, and this has actually not lead to an explicit 
construction of the Fock spaces.
In fact, despite their importance, an explicit construction of the parafermion or paraboson
Fock spaces was not known until recently (by an explicit construction is meant:
giving an complete orthogonal basis of the Fock space and the explicit action
of the parafermion or paraboson operators on these basis vectors). 
For the case of a finite set of parafermions, this explicit construction was 
given in~\cite{parafermion}, and for a finite set of parabosons in~\cite{paraboson}.

In the present paper, we extend the results of~\cite{paraboson,parafermion} to the case of an infinite
set of generators. 
This is not trivial, as the transition from finite rank algebras to infinite rank
algebras gives rise to some complications.
In section~2 of this paper, we define the parafermion and paraboson algebras
for an infinite set of generators, and specify the relation to 
$\so(\infty)$ and $\osp(1|\infty)$.
The main difficulty is however the construction of the Fock spaces $W(p)$ and $V(p)$.
In the case of a finite set of generators, a basis for the corresponding Fock spaces
was labelled by certain Gelfand-Zetlin (GZ) patterns.
Analysing the transformation of the basis under the action of generators,
we noticed that a certain stability takes place. 
It is this observation that allows us to extend our previously obtained results
to the infinite rank case.
For this purpose, it is convenient to define so-called stable infinite GZ-patterns,
introduced in section~3.
These are infinite triangular arrays of nonnegative integers, satisfying a ``betweenness
condition'' and containing only a finite number of distinct integers.
In section~4 we show that the parafermion algebra $\so(\infty)$ has an irreducible
action on the stable GZ-patterns with entries at most~$p$: these patterns
label the orthonormal basis vectors of the parafermion Fock space $W(p)$.
The proof that we are actually dealing with a representation relies on the
identities obtained in the finite rank case~\cite{parafermion}, and stability properties.
It is also interesting that the (formal) characters of these Fock spaces
$W(p)$ can be given, as a rather simple formula in terms of Schur functions~\eqref{charWp}.
Alternatively, the character can also be written as an expression including the
usual $\so(\infty)$ denominator.

In section~5 it is shown that the paraboson algebra $\osp(1|\infty)$ has an
irreducible action on the stable GZ-patterns with at most $p$ nonzero
entries per row. These patterns label the orthonormal basis of $V(p)$,
the paraboson Fock space. The result follows from the finite rank case~\cite{paraboson}
and certain stability properties of reduced matrix elements.
Once again, the characters of $V(p)$ have a simple expansion in terms of Schur
functions~\eqref{charVp}, or can be written as an expression displaying the
$\osp(1|\infty)$ denominator. 

To conclude, the computational techniques developed for the finite rank case in~\cite{paraboson,parafermion},
the (combinatorial) extension of certain finite GZ-patterns to stable infinite GZ-patterns,
and the stability properties of some reduced matrix elements obtained in~\cite{paraboson,parafermion}
allow us to offer a complete solution to the problem of the explicit construction
of parafermion and paraboson Fock representations.

\setcounter{equation}{0}
\section{The parafermion and paraboson algebras} \label{sec:algebras}

In this section we shall define a Lie algebra and a Lie superalgebra by means of a set of generators and
a set of relations. In both cases, the set of generators is an infinite set.
\begin{defi}
Let $\g=\so(\infty)$ be the (complex) Lie algebra with generators
\begin{equation}
\{ f_j^-, f_j^+ | j\in\Z_+=\{1,2,\ldots\}\;\}
\label{f-gens}
\end{equation}
and relations
\begin{equation}
[[f_{ j}^{\xi}, f_{ k}^{\eta}], f_{l}^{\epsilon}]=\frac 1 2
(\epsilon -\eta)^2
\delta_{kl} f_{j}^{\xi} -\frac 1 2  (\epsilon -\xi)^2
\delta_{jl}f_{k}^{\eta},  
\label{f-rels}
\end{equation}
where $j,k,l\in\Z_+$ and $\eta, \epsilon, \xi \in\{+,-\}$ (to be interpreted as $+1$ and $-1$
in the algebraic expressions $\epsilon -\xi$ and $\epsilon -\eta$).
\end{defi}
The relations~\eqref{f-rels} are also known as the {\em defining triple relations for a system of 
parafermions}~\cite{Green}, which is why $\g$ is often called the parafermion algebra. The reason for the notation
$\so(\infty)$ will be clear soon.

If one takes a finite set of such generators ($j\in\{1,2,\ldots,n\}$) and the corresponding
relations, the Lie algebra thus defined is the orthogonal Lie algebra $\so(2n+1)$, 
see~\cite{Ryan,Kamefuchi,parafermion}. 

Because of the defining relations~\eqref{f-rels}, it is easy to see that the following set is
a basis of~$\g$:
\begin{equation}
f_j^\pm \quad (j\in\Z_+), \qquad
[f_j^\pm, f_k^\pm] \quad(j < k;\;j,k\in\Z_+),\qquad 
[f_j^+, f_k^-] \quad( j,k\in\Z_+).
\label{so-basis}
\end{equation}
So, as a vector space, $\g$ consist of all finite linear combinations of the elements~\eqref{so-basis} with
complex coefficients. 

Consider now the elements
\begin{equation}
h_j = \frac{1}{2} [f_j^+,f_j^-] \quad (j\in\Z_+)
\label{hj}
\end{equation}
and the subspace $\h$ of $\g$ spanned by these elements. 
Using~\eqref{f-rels} it is easy to verify that all these elements commute, so $\h$ is an abelian
subalgebra of~$\g$. 
Furthermore, $\h$ is self-normalising. Hence $\h$ is a Cartan subalgebra of $\g$.

Let $\h^*$ be the vector space dual to $\h$, with dual basis $\{ \epsilon_j | j\in\Z_+\}$ defined by
\begin{equation}
\epsilon_j (h_k) = \delta_{jk}.
\end{equation}
All the elements from~\eqref{so-basis} are {\em root vectors} $x_\alpha$ of $\g$, in the sense that
\begin{equation}
[h_j, x_\alpha] = \alpha(h_j) x_\alpha,\qquad \forall j\in\Z_+,
\end{equation}
for some $\alpha\in\h^*$. The {\em roots} $\alpha$ corresponding to the elements~\eqref{so-basis}
are given by, respectively,
\begin{equation}
\pm\epsilon_j\ (j\in\Z_+), \qquad
\pm(\epsilon_j+\epsilon_k) \quad(j < k;\;j,k\in\Z_+),\qquad 
\epsilon_j-\epsilon_k \quad( j,k\in\Z_+).
\label{so-roots}
\end{equation}
The set of nonzero roots is thus
\begin{equation}
\Delta=\{ \pm\epsilon_j,\ \pm(\epsilon_j+\epsilon_k)\ (j < k), \ \epsilon_j-\epsilon_k \ (j\ne k)\}.
\label{so-Delta}
\end{equation}
So $\g$ has a root space decomposition, and one can verify (e.g.\ from the commutation relation
between root vectors) that $\g$ coincides with the Lie algebra~$\g'(B_\infty)$ as defined by 
Kac~\cite[\S 7.11]{Kac-LA}, sometimes also referred to as ``the Lie algebra $B_\infty$''.

For our purpose, it will be convenient to define the set of positive roots as
\begin{equation}
\Delta_+=\{ \epsilon_j,\ (\epsilon_j+\epsilon_k)\ (j < k), \ \epsilon_j-\epsilon_k \ (j<k)\}
\label{so-Delta+}
\end{equation}
and $\Delta_-$, the opposite set, as the set of negative roots. 
With $\n_+$ (resp.\ $\n_-$) defined as the subalgebra spanned by the positive (resp.\ negative)
root vectors, $\g$ has the usual triangular decomposition
\[
\g = \n_- \oplus \h \oplus \n_+.
\]
Note, however, that with the choice~\eqref{so-Delta+} it is not possible to give a set of simple roots.

\vskip 5mm
Now we shall define a closely related algebra~$\g$, which is however a Lie superalgebra~\cite{Kac} rather
than a Lie algebra. So this second algebra has a $\Z_2$-grading, $\g=\g_{\bar 0} \oplus \g_{\bar 1}$,
$\g_{\bar 0}$ being the even elements and $\g_{\bar 1}$ being the odd elements,
with the usual ``supercommutation relation'' (which is, in particular, an anticommutator $\{\cdot,\cdot\}$
when both elements are odd). 

\begin{defi}
Let $\g=\osp(1|\infty)$ be the (complex) Lie superalgebra with odd generators
\begin{equation}
\{ b_j^-, b_j^+ | j\in\Z_+\;\}
\label{b-gens}
\end{equation}
and relations
\begin{equation}
[\{ b_{ j}^{\xi}, b_{ k}^{\eta}\} , b_{l}^{\epsilon}]= (\epsilon -\xi) \delta_{jl} b_{k}^{\eta} 
 +  (\epsilon -\eta) \delta_{kl}b_{j}^{\xi}, 
\label{b-rels}
\end{equation}
where $j,k,l\in\Z_+$ and $\eta, \epsilon, \xi \in\{+,-\}$ (to be interpreted as $+1$ and $-1$
in the algebraic expressions $\epsilon -\xi$ and $\epsilon -\eta$).
\end{defi}
The relations~\eqref{b-rels} are the {\em defining triple relations for a system of 
parabosons}~\cite{Green}, and $\g$ is often called the paraboson algebra. 
For a finite set of such generators ($j\in\{1,2,\ldots,n\}$) and the corresponding
relations, the resulting Lie superalgebra is the orthosymplectic Lie superalgebra $\osp(1|2n)$, 
see~\cite{Ganchev,paraboson}. 

Note that this algebra could also be defined in an alternative way~\cite{KD}: one could also start
from the free algebra generated by the elements~\eqref{b-gens}, and consider the quotient with the
ideal generated by the elements corresponding to~\eqref{b-rels}. This construction gives the 
enveloping algebra $U(\g)$.

Once again, using the defining relations~\eqref{b-rels}, one finds a basis of~$\g$:
\begin{equation}
b_j^\pm \quad (j\in\Z_+), \qquad
\{b_j^\pm, b_k^\pm\} \quad(j < k;\;j,k\in\Z_+),\qquad 
\{b_j^+, b_k^-\} \quad( j,k\in\Z_+).
\label{osp-basis}
\end{equation}
In fact, $\g_{\bar 1}$ is spanned by the elements $b_j^\pm$ ($j\in\Z_+$), and
$\g_{\bar 0}$  by the remaining elements of~\eqref{osp-basis}.
Let us define the (even) elements
\begin{equation}
h_j = \frac{1}{2} \{b_j^+,b_j^-\} \quad (j\in\Z_+)
\label{hj-b}
\end{equation}
and the (even) subspace $\h$ of $\g$ spanned by these elements. 
One can again verify that $\h$ is a Cartan subalgebra of $\g$.
The dual vector space $\h^*$ is defined as before, with dual basis $\{ \epsilon_j | j\in\Z_+\}$.
All elements from~\eqref{osp-basis} are root vectors $x_\alpha$ of $\g$,
for some $\alpha\in\h^*$. 
The {\em roots} $\alpha$ corresponding to the elements~\eqref{osp-basis}
are given by the same set~\eqref{so-roots}. 
This time, the set of nonzero roots is $\Z_2$-graded, and one speaks of the
even roots $\Delta_0$ and the odd roots $\Delta_1$:
\begin{equation}
\Delta_1=\{ \pm\epsilon_j \}; \quad
\Delta_0=\{ \pm(\epsilon_j+\epsilon_k)\ (j < k), \ \epsilon_j-\epsilon_k \ (j\ne k)\}; \quad
\Delta=\Delta_0 \cup \Delta_1.
\label{osp-Delta}
\end{equation}
The set of positive roots is defined as before, and the odd and even positive roots are, resp.
\begin{equation}
\Delta_{1,+}=\{ \epsilon_j \},\quad
\Delta_{0,+}= \{ (\epsilon_j+\epsilon_k)\ (j < k), \ \epsilon_j-\epsilon_k \ (j<k)\}; \quad
\Delta_+= \Delta_{1,+} \cup \Delta_{0,+}.
\label{osp-Delta+}
\end{equation}
As before, $\Delta_-$, the opposite set, is the set of negative roots,
and $\g$ has a triangular decomposition.
This Lie superalgebra, which we have denoted by $\osp(1|\infty)$~\cite{Flato}, is sometimes also
denoted by~$B(0|\infty)$~\cite{Palev1992}.

In this section, we have given and specified the parafermion and paraboson algebras.
The main goal of the paper is the construction of their Fock representations.
For this purpose, it is convenient to introduce in the following section a number of (combinatorial)
quantities.

\setcounter{equation}{0}
\section{GZ-patterns, partitions and Schur functions} \label{sec:GZ}

Let us define infinite GZ-patterns, inspired by Gelfand-Zetlin patterns for 
$\u(n)$~\cite{GZ, Baird}, as follows.
\begin{defi}
A GZ-pattern $|m)$ is an infinite table or triangular array of nonnegative integers
$m_{ij}$ ($i,j\in\Z_+$, $i\leq j$), arranged as follows:
\begin{equation}
 |m) = \left|
\begin{array}{lcllll}
\vdots & \vdots & \cdots & \vdots & \qdots  \\
m_{1n} & m_{2n} & \cdots &  m_{nn}  &  \\
\vdots & \vdots & \qdots & & \\
m_{12} & m_{22} & & & \\
m_{11} & & & &
\end{array}
\right) 
= \left| \begin{array}{l} \vdots \\ {} [m]^n \\ \vdots \\ {} [m]^2 \\ {} [m]^1 \end{array} \right),
\label{GZ}
\end{equation}
such that the integer entries satisfy the following {\em betweenness conditions}
\begin{equation}
m_{i,j+1}\geq m_{ij}\geq m_{i+1,j+1}\qquad (\hbox{for all } i\leq j).
\label{between}
\end{equation}
\end{defi}
As already indicated in~\eqref{GZ}, we will denote the {\em rows} of $|m)$ as
$[m]^1=[m_{11}]$, $[m]^2=[m_{12},m_{22}]$, $[m]^3=[m_{13},m_{23},m_{33}]$, etc., and
refer to this as the first row, second row, third row etc.\ (we always count the rows
from bottom to top). So the $n$th row $[m]^n$
consists of a sequence of nonincreasing nonnegative integers of length~$n$.
Sometimes, it will be convenient to view $[m]^n$ as an infinite sequence
by adding zeros to the finite sequence:
\begin{equation}
[m]^n = [m_{1n},m_{2n},\ldots,m_{nn}] = [m_{1n},m_{2n},\ldots,m_{nn},0,0,\ldots ].
\label{mn}
\end{equation}

The infinite GZ-patterns that are relevant for us are those that stabilize after a number of rows,
counting from the bottom.
\begin{defi}
A GZ-pattern $|m)$ is stable if there exists a row index $N\in\Z_+$ such that 
\begin{equation}
[m]^n = [m]^N, \quad \hbox{for all } n>N.
\label{m=m}
\end{equation}
In such a case, one also says that $|m)$ is stable with respect to row~$N$.
\end{defi}
Note that, in order to speak of equal rows as in~\eqref{m=m}, one uses the extension by zeros as
in~\eqref{mn} (otherwise such rows would have unequal lengths).
For a stable GZ-pattern, it is not necessary to give an infinite table: giving just the first $N$
rows is sufficient. To indicate that one is still dealing with infinite GZ-patterns, 
we put an arrow above the entries of row~$N$. For example, the following GZ-pattern, which is
stable with respect to row~3, is denoted by:
\begin{equation}
\left| 
\begin{array}{l}\vdots\;\vdots \;\vdots\;\vdots\;\vdots\;\vdots\\[-1mm]
5 3 0 0 0\\[-1mm] 5 3 0 0 \\[-1mm] 5 3 0\\[-1mm]  3 1\\[-1mm] 2 \end{array} \right) =
 \left|
\begin{array}{l}
\uparrow  \uparrow  \uparrow \\[-1mm]
5  3  0 \\[-1mm]
3  1 \\[-1mm]
2 \end{array}
\right)
\label{530}
\end{equation}
Sometimes it will be useful to consider just the pattern consisting of the first $n$ rows of $|m)$ only.
We denote this by
\begin{equation}
 |m)^n = \left|
\begin{array}{lclll}
m_{1n} & m_{2n} & \cdots &  m_{nn}    \\
\vdots & \vdots & \qdots &  \\
m_{12} & m_{22} & &  \\
m_{11} & & & 
\end{array}
\right) 
\end{equation}
and shall refer to it as a $\u(n)$ GZ-pattern, as it coincides with the patterns in the common
GZ-basis of $\u(n)$~\cite{GZ, Baird}.

Let us also collect some notation on partitions and Schur functions, following Macdonald~\cite{Mac}.
A partition $\lambda$ is a (finite or infinite) sequence $\lambda=(\lambda_1,\lambda_2,\ldots)$ of
non-negative integers in decreasing order, $\lambda_1 \geq \lambda_2 \geq \cdots$, containing only
finitely many non-zero terms. The non-zero $\lambda_i$ are the parts of $\lambda$, and the number of
parts is the {\em length} $\ell(\lambda)$. The sum of the parts of $\lambda$ is
denoted by $|\lambda|=\lambda_1+\lambda_2+\cdots$.
The Young diagram of a partition $\lambda$ is the set of
left-adjusted rows of squares with $\lambda_i$ squares (or boxes) in the $i$th row, reading now from top
to bottom. For example, the Young diagram $F^\lambda$ of $\lambda = (5,4,4,1)$ is given by
\tabcolsep=2mm
\begin{equation}
F^\lambda = 
{\renewcommand{\arraystretch}{0.9}
\begin{tabular}{|c|c|c|c|c|}
\hline
 & & & & \\
\hline
 & & & \\
\cline{1-4}
 & & &  \\
\cline{1-4}
  \\
\cline{1-1}
\end{tabular} } .
\label{5441}
\end{equation}
The {\em conjugate} of $\lambda$ is the partition $\lambda'$ whose diagram is the transpose of the diagram
of $\lambda$ (i.e.\ by reflection in the main diagonal). For the above example, $\lambda'=(4,3,3,3,1)$. 
Another notation that will be useful is due to Frobenius. Let the main diagonal of $F^\lambda$
consist of $r$ boxes $(i,i)$ ($1\leq i \leq r$); let $\alpha_i=\lambda_i-i$ be the number of
boxes in the $i$th row to the right of $(i,i)$ and $\beta_i=\lambda_i'-i$ be the number of boxes
in the $i$th column below $(i,i)$. The Frobenius notation of $\lambda$ is
\begin{equation}
\lambda = \genfrac{(}{)}{0pt}{}{\alpha_1 \; \alpha_2 \cdots \alpha_r}{\beta_1\; \beta_2 \cdots \beta_r}.
\end{equation}
For example, the Frobenius notation of~\eqref{5441} is
\[
\genfrac{(}{)}{0pt}{}{4\; 2\; 1}{3\; 1\;0}.
\]

Partitions are used to label symmetric functions. One may consider symmetric polynomials 
in a finite number of variables $(x_1,x_2,\ldots,x_n)$, or the ring of symmetric functions~\cite{Mac}
in countably many variables $x=(x_1,x_2,\ldots )$. 
One particularly important set of symmetric functions are the Schur functions $s_\lambda$, labelled
by a partition~$\lambda$. For a finite number of variables (and $\ell(\lambda)\leq n$), the
Schur function (which is then a symmetric polynomial) is defined as a quotient of two
$(n\times n)$ determinants:
\begin{equation}
s_\lambda(x_1,\ldots,x_n) = \frac{\ds \det_{1\leq i,j\leq n} 
(x_i^{\lambda_j+n-j})}{\ds \det_{1\leq i,j\leq n} (x_i^{n-j})}.
\label{s_lambda}
\end{equation}
The number of variables can also be increased, and the Schur function $s_\lambda(x)$
for an infinite number of variables is well defined in the ring of symmetric functions~\cite{Mac}.

In the following section, we shall meet certain infinite series in the variables $x=(x_1,x_2,\ldots )$.
One such series, symmetric in all variables, is
\begin{equation}
E(x)= \prod_{i=1}^{\infty} (1-x_i) \prod_{1\leq i<j<\infty} (1-x_ix_j).
\label{Ex}
\end{equation}
This series has a nice expansion in terms of Schur functions~\cite[eq.~(11.9;5)]{Little}, namely
\begin{equation}
E(x) = \sum_{\eta\in{\cal E}} (-1)^{(|\eta|+r)/2} s_\eta(x),
\label{Exs}
\end{equation}
where the sum is over the set ${\cal E}$ of self-conjugate partitions $\eta$, which in the Frobenius
notation take the form
\begin{equation}
\eta = \genfrac{(}{)}{0pt}{}{\alpha_1 \; \alpha_2 \cdots \alpha_r}{\alpha_1\; \alpha_2 \cdots \alpha_r},
\end{equation}
and $|\eta|= 2(\alpha_1+\alpha_2+\cdots+\alpha_r)+r$. 
For example, the first terms of $E(x)$ read (with the ordinary $s_\lambda$ notation)
\begin{equation}
E(x)= 1 -s_{1}(x) + s_{21}(x) -s_{22}(x) -s_{311}(x) +s_{321}(x) +s_{4111}(x) -\cdots
\end{equation}
Two series of Schur functions that will play a role later are generalizations of~\eqref{Exs},
involving an extra positive integer~$p$.
The first is
\begin{equation}
E_{(p,0)}(x) = \sum_{\eta\in{\cal E}} (-1)^{(|\eta|+r)/2} s_{\eta_{(p,0)}}(x),
\label{Ep0}
\end{equation}
where for each $\eta\in{\cal E}$, the corresponding partition $\eta_{p,0}$ is defined by
the Frobenius form
\begin{equation}
\eta_{(p,0)} = \left( \begin{array}{cccc} \alpha_1+p & \alpha_2+p & \cdots & \alpha_r+p \\
 \alpha_1 & \alpha_2 & \cdots & \alpha_r \end{array}\right).
\end{equation}
The second one is
\begin{equation}
E_{(0,p)}(x) = \sum_{\eta\in{\cal E}} (-1)^{(|\eta|+r)/2} s_{\eta_{(0,p)}}(x),
\label{E0p}
\end{equation}
where for each $\eta\in{\cal E}$ the corresponding partition $\eta_{0,p}$ is defined by
\begin{equation}
\eta_{(0,p)} = \left( \begin{array}{cccc} \alpha_1 & \alpha_2 & \cdots & \alpha_r \\
 \alpha_1+p & \alpha_2+p & \cdots & \alpha_r+p \end{array}\right).
\end{equation}

\setcounter{equation}{0}
\section{The parafermion Fock spaces and $\so(\infty)$ representations} \label{sec:parafermion}

In~\cite{Greenberg}, the parafermion Fock space $W(p)$ of order~$p$ is defined, for any positive integer~$p$.
It is the Hilbert space with unique vacuum vector $|0\rangle$, 
defined by means of ($j,k=1,2,\ldots$)
\begin{align}
& \langle 0|0\rangle=1, \qquad f_j^- |0\rangle = 0, \qquad [f_j^-,f_k^+] |0\rangle =
 p\,\delta_{jk}\,|0\rangle,\label{pFock}\\
& (f_j^\pm)^\dagger = f_j^\mp,
\label{fdagger}
\end{align}
and by irreducibility under the action of the Lie algebra generated by 
the elements $f_j^+$, $f_j^-$ ($j=1,2,\ldots$), subject
to~\eqref{f-rels}, i.e.\ the Lie algebra~$\so(\infty)$. 
In~\eqref{fdagger}, $A^\dagger$ is the Hermitian adjoint of the operator $A$ with respect
to the inner product in the representation space $W(p)$. Following the common terminology,
such representations $W(p)$ are called unitary.

Note that for a finite number of parafermion generators $f_j^+$, $f_j^-$ ($j=1,2,\ldots,n$),
the corresponding Fock space $W(p)$ is a finite-dimensional irreducible unitary representation
of the Lie algebra $\so(2n+1)$. The structure of $W(p)$, and in particular the
parafermion operator actions in an appropriate orthogonal basis of $W(p)$, has
been determined in~\cite{parafermion} for the case of finite~$n$.

For the infinite rank case, $W(p)$ can also be defined as an induced module of the algebra
$\g=\so(\infty)$, introduced in Section~\ref{sec:algebras}.
First, note that the subalgebra of $\g$, spanned by the elements $[f_j^+,f_k^-]$ ($j,k\in\Z_+$),
is the infinite Lie algebra $\u(\infty)$ ($\g'(A_{+\infty})$ in the notation
of Kac~\cite{Kac-LA}, or $\gl(\infty)$ in the notation of~\cite{Palev1990}). 
Following~\cite{Palev1992,parafermion}, let us extend $\u(\infty)$ to a parabolic
subalgebra $\p$ of $\g$:
\begin{equation}
\p = \hbox{span} \{ [f_j^+, f_k^-] ;\ f_j^- ;\ 
[f_j^-, f_k^-] \ (j<k)  \} = \u(\infty) + \n_-.
\label{p}
\end{equation}
Since $[f_j^-,f_k^+] |0\rangle = p\,\delta_{jk}\, |0\rangle$ and $h_j=\frac{1}{2}[f_j^+,f_j^-]$,
one can consider the (one-dimensional) space spanned by $|0\rangle$ as the  
trivial one-dimensional $\u(n)$ module $\C |0\rangle$,
and refer to the {\em weight} of $|0\rangle$ as 
\begin{equation}
\wt(|0\rangle) =(-\frac{p}{2},-\frac{p}{2},\ldots).
\label{wt0}
\end{equation}
Since $f_j^- |0\rangle =0$,  the module $\C |0\rangle$ can be extended to a one-dimensional $\p$ module.
Then the Verma module or the induced $\g=\so(\infty)$ module $\overline W(p)$ is defined as:
\begin{equation}
 \overline W(p) = \hbox{Ind}_{\p}^{\g}\; \C|0\rangle.
 \label{defInd}
\end{equation}
This is an $\so(\infty)$ module with lowest weight $(-\frac{p}{2}, -\frac{p}{2},\ldots)$.
In general, $\overline W(p)$ is not an irreducible
representation of $\so(\infty)$. Let $M(p)$ be the maximal nontrivial submodule of $\overline W(p)$. Then the
simple module (irreducible representation), corresponding to the parafermion Fock space, is
\begin{equation}
W(p) = \overline W(p) / M(p).
\label{Wp}
\end{equation}

The main result of~\cite{parafermion} was, for the case of a finite number of
parafermion operators, the construction of an orthogonal basis for $W(p)$ and the action of these
operators on the basis elements. 
Here, we shall extend this to the infinite rank case.

We shall first describe the result, then state it in the form of a theorem, and end with the proof.
A basis of $W(p)$ consists of all stable GZ-patterns $|m)$ with $m_{ij}\leq p$ for all
$i$ and $j$. This basis is orthogonal:
\begin{equation}
( m' | m ) = \delta_{m,m'}
\label{inproduct}
\end{equation}
where the right hand side is 1 if $m_{ij}=m_{ij}'$ for all $(i,j)$ and 0 otherwise.
The action of the $\so(\infty)$ generators $f_j^\pm$ on the basis vectors $|m)$ is given by
\begin{align}
f_j^+|m) & = \sum_{m'} (m'|f_j^+|m) \; |m'),\label{f+}\\
f_j^-|m) & = \sum_{m'} (m'|f_j^-|m) \; |m'),\label{f-}
\end{align}
where these matrix elements are related by
\begin{equation}
(m|f_j^-|m') = (m'|f_j^+|m).
\label{f-f+}
\end{equation}
For a given $|m)$, there are only a finite number of non-zero matrix elements, so the sums in~\eqref{f+}
and~\eqref{f-} are finite. Let us describe the action of $f_j^+$ in~\eqref{f+}.
The only vectors appearing in the right hand side of~\eqref{f+} are (stable) GZ-patterns $|m')$ such that:
\begin{itemize}
\item For $n<j$, the rows of $|m)$ and $|m')$ are the same: $[m']^n=[m]^n$;
\item For $n\geq j$, the rows of $|m)$ and $|m')$ differ by~1 for one entry only.
More precisely, $[m']^n$ is the same sequence as $[m]^n$, apart from the fact that one entry 
(say, at position $\s{n}$) has been increased by~1. So for each $n\geq j$, there is a unique index 
denoted by $\s{n}$, with $\s{n}\in\{1,2,\ldots,n\}$, such that 
\begin{equation}
m_{\s{n},n}'=m_{\s{n},n}+1 \hbox{ and } m_{i,n}'=m_{i,n} \hbox{ for all } i\ne \s{n}.
\end{equation}
\end{itemize}
Since all GZ-patterns are stable, it follows from this rule that only a finite number of patterns $|m')$
appear in~\eqref{f+}.

The explicit expression of the matrix element $(m'|f_j^+|m)$ is deduced from that of
the finite rank case. 
Let us assume that $|m)$ is stable with respect to
row~$N$. 
First of all, for $j\leq N$, one has
\begin{equation}
\left( \begin{array}{l} \uparrow \\ {} [m']^N \\ \vdots \\ {} [m']^2 \\ {} [m']^1 \end{array} \right| f_j^+
\left| \begin{array}{l} \uparrow \\ {} [m]^N \\ \vdots \\ {} [m]^2 \\ {} [m]^1 \end{array} \right) =
\left(
\begin{array}{l} [m]^N  \\  \cdots \\ {[m]^j} \\ {[m]}^{j-1}\\ \cdots \\ m_{11} \end{array} \right. ;
\begin{array}{l} 100\cdots 0\\  \cdots \\ 10\cdots 0\\ 0\cdots 0 \\ \cdots \\ 0 \end{array} 
\left| 
\begin{array}{l} [m']^N  \\ \cdots \\ 
 {[m']^j} \\ {[m]}^{j-1}\\ \cdots \\ m_{11} \end{array} \right) 
G_{\s{N}} (m_{1N},m_{2N},\ldots,m_{NN}).
\label{cgc-G} 
\end{equation}
In the right hand side of~\eqref{cgc-G}, the first factor is a $\u(N)$ Clebsch-Gordan coefficient (CGC)~\cite{Klimyk}, of which
the expression will be given soon, and $G_k$ is the following function (see~\cite[Prop.~4]{parafermion}):
\begin{align}
& G_{k}(m_{1N}, m_{2N},\ldots,  m_{NN}) \nn\\
&
=\left(-\frac{
({\cal E}_N(m_{kN}+N-k)+1)\prod_{j\neq k=1}^{N} (m_{kN}-m_{jN}-k+j)}
{\prod_{j\neq  \frac{k}{2}=1}^{\lfloor N/2 \rfloor}  (m_{kN}-m_{2j,N}-k+2j)
(m_{kN}-m_{2j,N}-k+2j+1)}
\right)^{1/2} \quad(k\hbox{ even});  \label{Gkeven} \\
&
=\left(\frac{(p-m_{kN}+k-1)
({\cal O}_N(m_{kN}+N-k)+1)\prod_{j\neq k=1}^{N} (m_{kN}-m_{jN}-k+j)}
{\prod_{j\neq  \frac{k+1}{2}=1}^{\lceil N/2 \rceil}  (m_{kN}-m_{2j-1,N}-k+2j-1)
(m_{kN}-m_{2j-1,N}-k+2j)}
\right)^{1/2} \quad(k\hbox{ odd}); \label{Gkodd}
\end{align}
where $1\leq k\leq N$.
Herein ${\cal E}$ and ${\cal O}$ are the even and odd functions defined by
\begin{align}
& {\cal E}_{j}=1 \hbox{ if } j \hbox{ is even and 0 otherwise},\nn\\
& {\cal O}_{j}=1 \hbox{ if } j \hbox{ is odd and 0 otherwise}; \label{EO}
\end{align}
where obviously ${\cal O}_j=1-{\cal E}_j$, but it is convenient to use both notations.
Also, note that a product such as $\prod_{j\neq k=1}^{N}$ means ``the product over all $j$-values
running from 1 to $N$, but excluding $j=k$''. The notation $\lfloor a \rfloor$ (resp.\ $\lceil a \rceil$)
refers to the {\em floor} (resp.\ {\em ceiling}) of
$a$, i.e.\ the largest integer not exceeding~$a$ (resp.\ the smallest integer greater than or equal to $a$).

The $\u(N)$ CGC in~\eqref{cgc-G} is given by (see~\cite{Klimyk} or~\cite[Appendix~A]{paraboson})
\begin{align}
& \left(
\begin{array}{l} [m]^N  \\  \cdots \\ {[m]^j} \\ {[m]}^{j-1}\\ \cdots \\ m_{11} \end{array} \right. ;
\begin{array}{l} 100\cdots 0\\  \cdots \\ 10\cdots 0\\ 0\cdots 0 \\ \cdots \\ 0 \end{array} 
\left| 
\begin{array}{l} [m']^N  \\ \cdots \\ 
 {[m']^j} \\ {[m]}^{j-1}\\ \cdots \\ m_{11} \end{array} \right) = \nn\\
& S(\s{N},\s{N-1}) S(\s{N-1},\s{N-2})\ldots S(\s{j+1},\s{j})
\left( \frac{\prod_{k= 1}^{j-1}  
(l_{k,j-1}-l_{\s{j},j}-1 ) }
{ \prod_{k\neq \s{j}=1}^{j} (l_{kj}-l_{\s{j},j})}
\right)^{1/2}   \nn\\
&\times \prod_{r=j+1}^{N}
\left(  \prod_{k\ne \s{r-1}=1}^{r-1} \frac{(l_{k,r-1}-l_{\s{r},r}-1)}{(l_{k,r-1}-l_{\s{r-1},r-1}-1)}
\prod_{k\ne \s{r}=1}^{r} \frac{(l_{kr}-l_{\s{r-1},r-1})}{(l_{kr}-l_{\s{r},r})}
\right)^{1/2}  ,
\label{uN-CGC}
\end{align}
where, as usual,
\begin{equation}
l_{ij}=m_{ij}-i
\label{lij}
\end{equation}
and
\begin{equation}
S(k,r) = \left\{ \begin{array}{rcl}
 {1} & \hbox{for} & k\leq r  \\ 
 {-1} & \hbox{for} & k>r .
 \end{array}\right.
 \label{S}
\end{equation} 
Secondly, for $j>N$, only the rows with index $j$ or larger can change; one has
\begin{equation}
\left( \begin{array}{l} \uparrow \\ {}[m']^j\\ {}[m]^{j-1}{}\\ \vdots \\{}
 [m]^N \\ \vdots \\ {} [m]^2 \\ {} [m]^1 \end{array} \right| f_j^+
\left| \begin{array}{l} \\ \\ \\ \uparrow \\ {} [m]^N \\ \vdots \\ {} [m]^2 \\ {} [m]^1 \end{array} \right) =
\left( \frac{\prod_{k= 1}^{j-1}  
(l_{k,j-1}-l_{\s{j},j}-1 ) }
{ \prod_{k\neq \s{j}=1}^{j} (l_{kj}-l_{\s{j},j})}
\right)^{1/2}  
G_{\s{j}} (m_{1N},m_{2N},\ldots,m_{NN},0),
\label{cgc-G2} 
\end{equation}
where $\s{j}\in\{1,2,\ldots,N+1\}$.
Observe that, in fact, \eqref{cgc-G2} is a special case of~\eqref{cgc-G}. This can be seen by taking
$j$ as the stable index for $|m)$ in~\eqref{cgc-G2}: clearly, if $|m)$ is stable with respect to~$N$,
then it is also stable with respect to~$j$.

After this rather technical summary, let us state the main result.
\begin{theo}
The $\so(\infty)$ representation $W(p)$, i.e.\ the parafermion Fock representation of order~$p$,
has an orthonormal basis consisting of all stable GZ-patterns with entries at most~$p$.
The action of the $\so(\infty)$ generators $f_j^\pm$ on the basis vectors $|m)$
is given by~\eqref{f+}-\eqref{f-}, where the actual matrix elements are presented
in~\eqref{cgc-G} and~\eqref{cgc-G2}.
Under this action, $W(p)$ is a unitary irreducible $\so(\infty)$ representation,
generated by the vacuum vector $|0\rangle=|0)$ consisting of the GZ-pattern
with all zero entries.
\label{theo-f}
\end{theo}

\noindent {\bf Proof}. The proof goes in a number of steps, the main part relying heavily
on the corresponding result for a finite number of parafermion operators.
First of all, orthonormality is imposed by~\eqref{inproduct}.
The hardest part is to prove that under the given action $W(p)$ is actually
a representation of $\so(\infty)$. For this, it is sufficient to prove that
\begin{equation}
(m'| A_{jkl}^{\xi\eta\epsilon} |m) = 0, 
\label{suff}
\end{equation}
for all couples of basis vectors of $W(p)$ and for all 
\begin{equation}
A_{jkl}^{\xi\eta\epsilon} =[[f_{ j}^{\xi}, f_{ k}^{\eta}], f_{l}^{\epsilon}]-\frac 1 2
(\epsilon -\eta)^2 \delta_{kl} f_{j}^{\xi} + \frac 1 2  (\epsilon -\xi)^2
\delta_{jl}f_{k}^{\eta},  
\end{equation}
since these expressions correspond to the defining relations~\eqref{f-rels} of $\so(\infty)$.
Let $|m)$ be stable with respect to some value~$n$, let $|m')$ be stable with respect to some value~$n'$,
and for a given $A_{jkl}^{\xi\eta\epsilon}$, define
\[
N=\max(n,n',j,k,l).
\]
Using~\cite[Theorem~6]{parafermion}, it follows that 
\begin{equation}
\left( \begin{array}{l} [m']^N \\ \vdots \\ {} [m']^2 \\ {} [m']^1 \end{array} \right| 
A_{jkl}^{\xi\eta\epsilon}
\left| \begin{array}{l} [m]^N \\ \vdots \\ {} [m]^2 \\ {} [m]^1 \end{array} \right) = 0
\end{equation}
as an identity in $\so(2N+1)$. Hence, by stability,
\begin{equation}
\left( \begin{array}{l} \uparrow \\ {} [m']^N \\ \vdots \\ {} [m']^2 \\ {} [m']^1 \end{array} \right| 
A_{jkl}^{\xi\eta\epsilon}
\left| \begin{array}{l} \uparrow \\ {} [m]^N \\ \vdots \\ {} [m]^2 \\ {} [m]^1 \end{array} \right)
= (m'| A_{jkl}^{\xi\eta\epsilon} |m) = 0.
\end{equation}
So we are dealing with a representation.
The representation is unitary, since the matrix elements of $f_j^-$ are defined by means of those
of $f_j^+$ and~\eqref{f-f+}.
The basis vector $|0)$ (zero GZ-pattern) of $W(p)$ satisfies
\[
(0|0)=1, \qquad f_j^- |0) = 0, \qquad [f_j^-,f_k^+] |0) = f_j^- f_k^+ |0) =
 p\,\delta_{jk}\,|0)
\]
so it is the vacuum vector $|0\rangle$. $W(p)$ is generated by $|0)$: 
indeed, let $|m)$ be any basis vector of $W(p)$. Suppose $|m)$
is stable with respect to row~$N$. Consider the corresponding $\u(N)$
GZ-pattern consisting of the first $N$ rows of $|m)$. This is a vector
of the unitary irreducible $\so(2N+1)$ representation described in~\cite[Theorem~6]{parafermion}.
Hence there exists an element $A$ from $\so(2N+1)$, i.e.\ $A$ can be expressed as a 
linear combination of products of the generators $f^\pm_j$ with $j\in\{1,2,\ldots,N\}$,
such that
\[
\left| \begin{array}{l} [m]^N \\ \vdots \\ {} [m]^2 \\ {} [m]^1 \end{array} \right) =
A\; \left| \begin{array}{l} [0]^N \\ \vdots \\ {} [0]^2 \\ {} [0]^1 \end{array} \right).
\]
Hence, for this same expression $A$,
\[
|m) = A\, |0).
\]
This also yields irreducibility, since $(m|A|0)=1$ implies $(0|A^\dagger|m)=1$. 
Due to the fact that all entries in such a $\u(N)$ GZ-pattern are at most~$p$,
this property holds also for the basis vectors of $W(p)$.  \mybox

\vskip 3mm
The matrix elements in~\eqref{cgc-G} look complicated at first sight, but in fact they are
not so difficult to use explictly. Let us give an example, and compute the action
of $f_1^+, f_2^+, \ldots$ on the vector~\eqref{530} of~$W(p)$ (where, obviously, $p\geq 5$).
One finds:
\begin{align}
f_1^+  \left|
\begin{array}{l}
\uparrow  \uparrow  \uparrow \\[-1mm]
5  3  0 \\[-1mm]
3  1 \\[-1mm]
2 \end{array}
\right) & = \sqrt{\frac{5(p-5)}{7}}
\left|
\begin{array}{l}
\uparrow  \uparrow  \uparrow \\[-1mm]
6  3  0 \\[-1mm]
4  1 \\[-1mm]
3 \end{array}
\right)  - \sqrt{5}
\left|
\begin{array}{l}
\uparrow  \uparrow  \uparrow \\[-1mm]
5  4  0 \\[-1mm]
4  1 \\[-1mm]
3 \end{array} 
\right) - \frac{1}{3}\sqrt{\frac{p+2}{14}}
\left|
\begin{array}{l}
\uparrow  \uparrow  \uparrow \\[-1mm]
5  3  1 \\[-1mm]
4  1 \\[-1mm]
3 \end{array}
\right) \nn\\
& - \sqrt{\frac{2(p-5)}{7}} 
\left|
\begin{array}{l}
\uparrow  \uparrow  \uparrow \\[-1mm]
6  3  0 \\[-1mm]
3  2 \\[-1mm]
3 \end{array}
\right) + \frac{1}{3}\sqrt{\frac{10(p+2)}{7}}
\left|
\begin{array}{l}
\uparrow  \uparrow  \uparrow \\[-1mm]
5  3  1 \\[-1mm]
3  2 \\[-1mm]
3 \end{array}
\right) \nn\\
f_2^+  \left|
\begin{array}{l}
\uparrow  \uparrow  \uparrow \\[-1mm]
5  3  0 \\[-1mm]
3  1 \\[-1mm]
2 \end{array}
\right) & = \sqrt{\frac{5(p-5)}{7}}
\left|
\begin{array}{l}
\uparrow  \uparrow  \uparrow \\[-1mm]
6  3  0 \\[-1mm]
4  1 \\[-1mm]
2 \end{array}
\right) -\sqrt{5}
\left|
\begin{array}{l}
\uparrow  \uparrow  \uparrow \\[-1mm]
5  4  0 \\[-1mm]
4  1 \\[-1mm]
2 \end{array} 
\right) -\frac{1}{3}\sqrt{\frac{p+2}{14}}
\left|
\begin{array}{l}
\uparrow  \uparrow  \uparrow \\[-1mm]
5  3  1 \\[-1mm]
4  1 \\[-1mm]
2 \end{array}
\right) \nn\\
& +  \sqrt{\frac{2(p-5)}{7}}
\left|
\begin{array}{l}
\uparrow  \uparrow  \uparrow \\[-1mm]
6  3  0 \\[-1mm]
3  2 \\[-1mm]
2 \end{array}
\right) -\frac{1}{3}\sqrt{\frac{10(p+2)}{7}}
\left|
\begin{array}{l}
\uparrow  \uparrow  \uparrow \\[-1mm]
5  3  1 \\[-1mm]
3  2 \\[-1mm]
2 \end{array}
\right) \nn\\
f_3^+  \left|
\begin{array}{l}
\uparrow  \uparrow  \uparrow \\[-1mm]
5  3  0 \\[-1mm]
3  1 \\[-1mm]
2 \end{array}
\right) & = \sqrt{\frac{18(p-5)}{7}}
\left|
\begin{array}{l}
\uparrow  \uparrow  \uparrow \\[-1mm]
6  3  0 \\[-1mm]
3  1 \\[-1mm]
2 \end{array}
\right)  + \sqrt{\frac{2(p+2)}{21}}
\left|
\begin{array}{l}
\uparrow  \uparrow  \uparrow \\[-1mm]
5  3  1 \\[-1mm]
3  1 \\[-1mm]
2 \end{array}
\right) \nn\\
f_4^+  \left|
\begin{array}{l}
\uparrow  \uparrow  \uparrow  \\[-1mm]
5  3  0 \\[-1mm]
3  1 \\[-1mm]
2 \end{array}
\right) & = \sqrt{\frac{4(p-5)}{7}}
\left|
\begin{array}{l}
\uparrow  \uparrow  \uparrow \uparrow \\[-1mm]
6  3  0  0 \\[-1mm]
5  3  0 \\[-1mm]
3  1 \\[-1mm]
2 \end{array}
\right) + \sqrt{2}
\left|
\begin{array}{l}
\uparrow  \uparrow  \uparrow \uparrow \\[-1mm]
5  4  0  0 \\[-1mm]
5  3  0 \\[-1mm]
3  1 \\[-1mm]
2 \end{array} 
\right) + \sqrt{\frac{3(p+2)}{7}}
\left|
\begin{array}{l}
\uparrow  \uparrow  \uparrow \uparrow \\[-1mm]
5  3  1  0 \\[-1mm]
5  3  0 \\[-1mm]
3  1 \\[-1mm]
2 \end{array}
\right) \nn
\end{align}
Note that for $p=5$ some matrix elements become automatically zero, corresponding to the
coefficients in front of vectors that do not belong to $W(5)$.

\vskip 3mm
As a second example, consider the case $p=1$, which should reproduce the known fermion Fock space.
For $p=1$, the entries in the GZ-patterns of $W(1)$ are~0 or~1. It is easy to compute that
\begin{equation}
f_j^+ |0) = 
\left| \begin{array}{l} \uparrow \\{} [1]^j \\{}[0]^{j-1}\\ \vdots \\  {} [0]^1 \end{array} \right)
\end{equation}
and therefore
\[
f_j^+ f_j^+ |0) = 0.
\]
Furthermore,
\begin{equation}
f_j^+ f_k^+ |0) = f_j^+ 
\left| \begin{array}{l} \uparrow \\{} [1]^k \\{}[0]^{k-1}\\ \vdots \\  {} [0]^1 \end{array} \right)
=- \left| \begin{array}{l} \uparrow \\{} [11]^k \\{}[1]^{k-1}\\ \vdots \\{}[1]^j\\{}[0]^{j-1}\\ \vdots \\  {} [0]^1 \end{array} \right)
\qquad \hbox{for } j<k,
\end{equation}
and $f_k^+ f_j^+ |0)= - f_j^+ f_k^+ |0)$. 
These configurations generalize, and since the entries in the GZ-patterns consist of zeros and ones only, 
one can write them in a more appropriate form:
\begin{equation}
|m) = v(\theta_1,\theta_2,\theta_3,\ldots), \qquad
\theta_n = \sum_{i} m_{in} - \sum_i m_{i,n-1},
\end{equation}
where $m_{i0}=0$.
So each $\theta_n\in\{0,1\}$, and $(\theta_1,\theta_2,\theta_3,\ldots)$ is an infinite string of zeros and ones
with only a finite number of ones, because of the stability of $|m)$.
It is not difficult to see that 
\begin{equation}
f_{i_1}^+ f_{i_2}^+ \cdots f_{i_n}^+ |0) = (-1)^{n-1} v(0,\ldots,0,1,0,\ldots,0,1,0,\ldots),
\quad (i_1<i_2<\cdots<i_n),
\end{equation}
where the string has ones at positions $i_1,i_2,\ldots, i_n$. 
Hence the $v$-basis, or equivalently the $|m)$-basis, coincides (up to a sign)
with the usual fermion Fock basis when $p=1$.

\vskip 3mm
Let us now return to the general case, with $p$ arbitrary.
Recall that the weight of $|0)$ is given by~\eqref{wt0}. 
In other words, $|0)$ is $\h$-diagonal, and
\begin{equation}
h_j |0) = -\frac{p}{2} |0).
\end{equation}
Using $h_j=\frac{1}{2}[f_j^+,f_j^-]$ and the triple relations~\eqref{f-rels}, one
finds $[h_j, f_k^\pm] = \pm \delta_{jk} f_k^\pm$. 
Since all vectors of $W(p)$ are generated from $|0)$ by acting with $f_k^\pm$ ($k=1,2,\ldots$),
it follows that each basis vector $|m)$ is $\h$-diagonal, and that
\begin{equation}
h_j |m) = ( -\frac{p}{2} + \sum_i m_{i,j} - \sum_i m_{i,j-1} ) |m).
\end{equation}
The sequence of eigenvalues of $(h_1,h_2,h_3,\ldots)$ is the weight of $|m)$:
\begin{equation}
\wt(|m)) = (k_1,k_2,k_3,\ldots), \qquad k_j= -\frac{p}{2} + \sum_i m_{i,j} - \sum_i m_{i,j-1}.
\end{equation}
The character of $W(p)$ is the expression
\begin{equation}
\ch W(p) = \sum_{ |m) \in W(p) } e^{\wt(|m))},
\end{equation}
where $e$ is the formal exponential. 
It is common to write, in this context,
\begin{equation}
e^{\wt(|m))} = x_1^{k_1} x_2^{k_2} x_3^{k_3} \cdots ,
\end{equation}
so
\begin{equation}
\ch W(p) = \sum_{ |m) \in W(p) } x_1^{k_1} x_2^{k_2} x_3^{k_3} \cdots
\end{equation}
and it becomes a (symmetric) function in the variables $x_1, x_2, \ldots$.
Then we have
\begin{theo}
The character of $W(p)$ is given by
\begin{equation}
\ch W(p) = (x_1x_2\cdots)^{-p/2} \sum_{\lambda,\; \ell(\lambda')\leq p} s_\lambda (x),
\label{charWp}
\end{equation}
where the sum is over all partitions $\lambda$ with largest part not exceeding~$p$.
\end{theo}
For example,
\begin{align}
\ch W(1) &= (x_1x_2\cdots)^{-1/2} (1+s_{1}(x)+s_{11}(x)+s_{111}(x)+\cdots),\nn\\
\ch W(2) &= (x_1x_2\cdots)^{-1} (1+s_{1}(x)+s_2(x)+s_{11}(x)+s_{21}(x)+s_{111}(x)+\cdots).\nn
\end{align}

The proof of~\eqref{charWp} follows from the finite rank case~\cite{parafermion}. 
Let $N$ be a fixed positive number, and consider all basis vectors $|m)$ in $W(p)$ that
are stable with respect to row~$N$. By~\cite[Corollary~5]{parafermion}, one has
\begin{equation}
\sum_{\ds \genfrac{}{}{0pt}{}{\scriptstyle |m) \in W(p)}{\scriptstyle |m){\scriptsize\hbox{ stable w.r.t. row }}N} } e^{\wt(|m))} = 
(x_1x_2\cdots x_N)^{-p/2} \sum_{\lambda,\; \ell(\lambda')\leq p} s_\lambda (x_1,x_2,\ldots,x_N).
\end{equation}
The assertion then follows from a limit process.

The character can be written in an alternative form. This is based on an identity
for Schur functions with a finite number of variables. 
From~\cite[p.~84, eq.~($2^\prime$)]{Mac}, one finds
\[
\sum_{\ell(\lambda')\leq p} s_\lambda(x_1,\ldots,x_n) = \frac{\det ( x_i^{p+2n-j}-x_i^{j-1})}{\det ( x_i^{2n-j}-x_i^{j-1})},
\]
and using this last expression, King~\cite{King} has shown:
\[
\sum_{\ell(\lambda')\leq p} s_\lambda(x_1,\ldots,x_n) = \frac{E_{(p,0)}(x_1,\ldots,x_n)}{E(x_1,\ldots,x_n)} =
\frac{E_{(p,0)}(x_1,\ldots,x_n)}{\prod_i (1-x_i)\prod_{i<j}(1-x_ix_j)},
\]
where $E$ and $E_{(p,0)}$ are the expressions given in~\eqref{Ex} and~\eqref{Ep0}, but restricted to
a finite number of variables.
Hence we can write
\begin{align}
\ch W(p) &= (x_1x_2\cdots)^{-p/2} \frac{E_{(p,0)}(x)}{\prod_i (1-x_i)\prod_{i<j}(1-x_ix_j)} \nn\\
& = (x_1x_2\cdots)^{-p/2}
\frac{\sum_{\eta\in{\cal E}} (-1)^{(|\eta|+r)/2} s_{\eta_{(p,0)}}(x)}{\prod_i (1-x_i)\prod_{i<j}(1-x_ix_j)},
\label{charWp-den}
\end{align}
where the notation follows that of Section~\ref{sec:GZ}.
Note that~\eqref{charWp} gives the character of $W(p)$ as an expansion in Schur functions, 
whereas~\eqref{charWp-den} is of a different type. In~\eqref{charWp-den}, the denominator
factors correspond to the positive roots $\epsilon_i$ and $\epsilon_i+\epsilon_j$ ($i<j$) of
$\so(\infty)$ (the factors corresponding to the positive roots $\epsilon_i+\epsilon_j$ ($i<j$)
could be considered as part of the Schur functions $s_{\eta_{(p,0)}}(x)$ in the numerator).
So~\eqref{charWp-den} can be regarded as a character formula written by means of the common
$\so(\infty)$ denominator $\prod_{\alpha\in\Delta_+}(e^{\alpha/2}-e^{-\alpha/2})$. 

\setcounter{equation}{0}
\section{The paraboson Fock spaces and $\osp(1|\infty)$ representations} \label{sec:paraboson}

The paraboson Fock space $V(p)$ of order~$p$, with~$p$ any positive integer~$p$,
is the Hilbert space with unique vacuum vector $|0\rangle$, 
defined by means of ($j,k=1,2,\ldots$)~\cite{Greenberg}
\begin{align}
& \langle 0|0\rangle=1, \qquad b_j^- |0\rangle = 0, \qquad \{b_j^-,b_k^+\} |0\rangle =
 p\,\delta_{jk}\,|0\rangle,\label{pFock-b}\\
& (b_j^\pm)^\dagger = b_j^\mp,
\label{bdagger}
\end{align}
and by irreducibility under the action of the Lie superalgebra generated by 
the elements $b_j^+$, $b_j^-$ ($j=1,2,\ldots$), subject
to~\eqref{b-rels}, i.e.\ the Lie superalgebra~$\osp(1|\infty)$. 
For a finite number of paraboson operators $b_j^+$, $b_j^-$ ($j=1,2,\ldots,n$),
the corresponding Fock space $V(p)$ is an infinite dimensional irreducible unitary representation
of the Lie superalgebra $\osp(1|2n)$, which has been studied in detail in~\cite{paraboson}.

For the infinite rank case, one can define $V(p)$ as an induced module of $\g=\osp(1|\infty)$,
just like in the previous section.
The subalgebra of $\g$ spanned by the elements $\{b_j^+,b_k^-\}$ ($j,k\in\Z_+$)
is again the infinite Lie algebra $\u(\infty)$. 
One extends $\u(\infty)$ to a parabolic subalgebra $\p$ of $\g$:
\begin{equation}
\p = \hbox{span} \{ \{b_j^+, b_k^-\} ;\ b_j^- ;\ 
\{b_j^-, b_k^-\} \ (j<k)  \} = \u(\infty) + \n_-.
\label{p-b}
\end{equation}
Using $\{b_j^-,b_k^+\} |0\rangle = p\,\delta_{jk}\, |0\rangle$ and $h_j=\frac{1}{2}\{b_j^+,b_j^-\}$,
the (one-dimensional) space spanned by $|0\rangle$ is a  
trivial one-dimensional $\u(n)$ module $\C |0\rangle$,
with weight 
\begin{equation}
\wt(|0\rangle) =(\frac{p}{2},\frac{p}{2},\ldots).
\label{wt0-b}
\end{equation}
As $|0\rangle$ is annihilated by all $b_j^-$,
$\C |0\rangle$ can be extended to a one-dimensional $\p$ module.
Then the induced $\g=\osp(1|\infty)$ module $\overline V(p)$ is defined as:
\begin{equation}
 \overline V(p) = \hbox{Ind}_{\p}^{\g}\; \C|0\rangle.
 \label{defIndV}
\end{equation}
This is an $\osp(1|\infty)$ module with lowest weight $(\frac{p}{2}, \frac{p}{2},\ldots)$;
the corresponding simple module (irreducible representation) is obtained by taking the
quotient with the maximal nontrivial submodule $M(p)$:
\begin{equation}
V(p) = \overline V(p) / M(p).
\label{Vp}
\end{equation}

In~\cite{paraboson} we obtained, for the case of a finite number of
paraboson operators, an orthonormal basis for $V(p)$ and the action of these
operators on the basis elements. Now this will be extended to the infinite case.
As the outcome and methods are similar to those of the previous section, 
we shall only describe the main results. 
A basis of $V(p)$ consists of all stable GZ-patterns $|m)$ with 
\begin{equation}
m_{ij}=0 \hbox{ whenever }i>p. 
\label{pbcondition}
\end{equation}
In other words, the GZ-patterns have only $p$ columns, the rest of the columns consisting
of zeroes only.
This basis is orthogonal.
The action of the $\osp(1|\infty)$ generators $b_j^\pm$ on the basis vectors $|m)$ is given by
\begin{align}
b_j^+|m) & = \sum_{m'} (m'|b_j^+|m) \; |m'),\label{b+}\\
b_j^-|m) & = \sum_{m'} (m'|b_j^-|m) \; |m'),\label{b-}
\end{align}
where these matrix elements are related by
\begin{equation}
(m|b_j^-|m') = (m'|b_j^+|m).
\label{b-b+}
\end{equation}
The sums in~\eqref{b+} and~\eqref{b-} are again finite. 
The action of $b_j^+$ in~\eqref{b+} is easy to describe: 
the only vectors appearing in the right hand side of~\eqref{b+} are (stable) GZ-patterns $|m')$ such that:
\begin{itemize}
\item For $n<j$, the rows of $|m)$ and $|m')$ are the same: $[m']^n=[m]^n$;
\item For $n\geq j$, the rows of $|m)$ and $|m')$ differ by~1 for only one entry.
More precisely, $[m']^n$ is the same sequence as $[m]^n$, apart from the fact that one entry 
(say, at position $\s{n}$) has been increased by~1. So for each $n\geq j$, there is a unique index 
denoted by $\s{n}$, with $\s{n}\in\{1,2,\ldots,p\}$, such that 
\begin{equation}
m_{\s{n},n}'=m_{\s{n},n}+1 \hbox{ and } m_{i,n}'=m_{i,n} \hbox{ for all } i\ne \s{n}.
\end{equation}
\end{itemize}

As in the previous section, the main ingredient is the explicit expression 
of the matrix element $(m'|b_j^+|m)$, deduced from that of the finite rank case. 
Assume that $|m)$ is stable with respect to row~$N$. 
It is again appropriate to make a distinction between $j\leq N$ and $j> N$.
For $j\leq N$, one has
\begin{equation}
\left( \begin{array}{l} \uparrow \\ {} [m']^N \\ \vdots \\ {} [m']^2 \\ {} [m']^1 \end{array} \right| b_j^+
\left| \begin{array}{l} \uparrow \\ {} [m]^N \\ \vdots \\ {} [m]^2 \\ {} [m]^1 \end{array} \right) =
\left(
\begin{array}{l} [m]^N  \\  \cdots \\ {[m]^j} \\ {[m]}^{j-1}\\ \cdots \\ m_{11} \end{array} \right. ;
\begin{array}{l} 100\cdots 0\\  \cdots \\ 10\cdots 0\\ 0\cdots 0 \\ \cdots \\ 0 \end{array} 
\left| 
\begin{array}{l} [m']^N  \\ \cdots \\ 
 {[m']^j} \\ {[m]}^{j-1}\\ \cdots \\ m_{11} \end{array} \right) 
\tilde F_{\s{N}} (m_{1N},m_{2N},\ldots,m_{pN}).
\label{cgc-F} 
\end{equation}
Herein, the first factor of the right hand side is the same $\u(N)$ Clebsch-Gordan coefficient~\cite{Klimyk}
as before, see~\eqref{uN-CGC}.
The second function, with $\s{N}\in\{1,2,\ldots,p\}$ because of~\eqref{pbcondition},
can be deduced from~\cite[Prop.~6]{paraboson}:
\begin{align}
\tilde F_{k}(m_{1N}, m_{2N},\ldots,  m_{pN})  = &
(-1)^{m_{k+1,N}+\cdots+m_{pN}} (m_{kN}+p+1-k)^{1/2} \nn\\
& \times  \prod_{j\neq k=1}^{p} \left( \frac{m_{jN}-m_{kN}-j+k }{m_{jN}-m_{kN}-j+k-{\cal O}_{m_{jN}-m_{kN}} }
\right)^{1/2}. \label{Fk}
\end{align}
Herein ${\cal O}$ has been defined before~\eqref{EO}.

For $j>N$, only the rows with index $j$ or larger can change, and the result is simply
\begin{equation}
\left( \begin{array}{l} \uparrow \\ {}[m']^j\\ {}[m]^{j-1}{}\\ \vdots \\{}
 [m]^N \\ \vdots \\ {} [m]^2 \\ {} [m]^1 \end{array} \right| b_j^+
\left| \begin{array}{l} \\ \\ \\ \uparrow \\ {} [m]^N \\ \vdots \\ {} [m]^2 \\ {} [m]^1 \end{array} \right) =
\left( \frac{\prod_{k= 1}^{j-1}  
(l_{k,j-1}-l_{\s{j},j}-1 ) }
{ \prod_{k\neq \s{j}=1}^{j} (l_{kj}-l_{\s{j},j})}
\right)^{1/2}  
\tilde F_{\s{j}} (m_{1N},m_{2N},\ldots,m_{pN}),
\label{cgc-F2} 
\end{equation}
where again $\s{j}\in\{1,2,\ldots,p\}$. As before, 
\eqref{cgc-F2} can be seen as a special case of~\eqref{cgc-F}. 

We can now state:
\begin{theo}
The $\osp(1|\infty)$ representation $V(p)$, i.e.\ the paraboson Fock representation of order~$p$,
has an orthonormal basis consisting of all stable GZ-patterns with at most~$p$ nonzero
entries per row.
The action of the $\osp(1|\infty)$ generators $b_j^\pm$ on the basis vectors $|m)$
is given by~\eqref{b+}-\eqref{b-}, where the actual matrix elements are presented
in~\eqref{cgc-F} and~\eqref{cgc-F2}.
Under this action, $V(p)$ is a unitary irreducible $\osp(1|\infty)$ representation,
generated by the vacuum vector $|0\rangle=|0)$ consisting of the GZ-pattern
with all zero entries.
\label{theo-b}
\end{theo}

The proof is essentially the same as that of Theorem~\ref{theo-f}. The essential
part is again to show that under the given actions $V(p)$ is a representation of~$\osp(1|\infty)$.
Here, one uses the results of~\cite{paraboson} for the case of a finite number of paraboson operators.
Note that, using~\cite[Prop.~6]{paraboson}, one would find instead of~\eqref{Fk}:
\begin{align}
F_{k}(m_{1N}, m_{2N},\ldots,  m_{pN},0,\ldots,0)  = &
(-1)^{m_{k+1,N}+\cdots+m_{pN}} (m_{kN}+N+1-k+ {\cal E}_{m_{kN}}(p-N))^{1/2} \nn\\
& \times  \prod_{j\neq k=1}^{N} \left( \frac{m_{jN}-m_{kN}-j+k }{m_{jN}-m_{kN}-j+k-{\cal O}_{m_{jN}-m_{kN}} }
\right)^{1/2}, 
\label{FF}
\end{align}
with $k\in\{1,2,\ldots,p\}$. 
When $m_{kN}$ is even, \eqref{FF} immediately reduces to~\eqref{Fk} (the product $j\neq k=1$ upto $N$
becomes a product upto $p$ as $m_{jN}=0$ for all $j>p$).
When $m_{kN}$ is odd, the separate first factor in~\eqref{FF} becomes $(m_{kN}+N+1-k)^{1/2}$. However, writing
out explicitly all remaining factors in the product~\eqref{FF} (and using $m_{jN}=0$ for all $j>p$) many of these 
cancel because they appear in the numerator and in the denominator, and also the separate factor
cancels with one in the denominator. The remaining expression is again~\eqref{Fk}.

We also mention, without giving further details, that the case $p=1$ yields the
known boson Fock space.

As a final result, let us give the character of $V(p)$.
The generating vector $|0)$ is $\h$-diagonal, and
\begin{equation}
h_j |0) = \frac{p}{2} |0).
\end{equation}
Here, $h_j=\frac{1}{2}\{b_j^+,b_j^-\}$ and using the triple relations~\eqref{b-rels} one
finds $[h_j, b_k^\pm] = \pm \delta_{jk} b_k^\pm$. 
It follows that each basis vector $|m)$ is $\h$-diagonal, with
\begin{equation}
h_j |m) = ( \frac{p}{2} + \sum_i m_{i,j} - \sum_i m_{i,j-1} ) |m),
\end{equation}
so the weight of $|m)$ is 
\begin{equation}
\wt(|m)) = (k_1,k_2,k_3,\ldots), \qquad k_j= \frac{p}{2} + \sum_i m_{i,j} - \sum_i m_{i,j-1}.
\end{equation}
The character of $V(p)$ is 
\begin{equation}
\ch V(p) = \sum_{ |m) \in V(p) } x_1^{k_1} x_2^{k_2} x_3^{k_3} \cdots
\end{equation}
and we have
\begin{theo}
The character of $V(p)$ is given by
\begin{equation}
\ch V(p) = (x_1x_2\cdots)^{p/2} \sum_{\lambda,\; \ell(\lambda)\leq p} s_\lambda (x),
\label{charVp}
\end{equation}
where the sum is over all partitions $\lambda$ of length at most~$p$.
\end{theo}
For example,
\begin{align}
\ch V(1) &= (x_1x_2\cdots)^{1/2} (1+s_{1}(x)+s_{2}(x)+s_{3}(x)+\cdots),\nn\\
\ch V(2) &= (x_1x_2\cdots) (1+s_{1}(x)+s_2(x)+s_{11}(x)+s_{3}(x)+s_{21}(x)+\cdots).\nn
\end{align}

Also here, the character can be written in an alternative form. 
For Schur functions with a finite number of variables,
one has~\cite{King} (or~\cite[Eq.~(5.17)]{paraboson})
\[
\sum_{\ell(\lambda)\leq p} s_\lambda(x_1,\ldots,x_n) = \frac{E_{(0,p)}(x_1,\ldots,x_n)}{E(x_1,\ldots,x_n)} =
\frac{E_{(0,p)}(x_1,\ldots,x_n)}{\prod_i (1-x_i)\prod_{i<j}(1-x_ix_j)},
\]
where $E$ and $E_{(0,p)}$ are the expressions given in~\eqref{Ex} and~\eqref{E0p}, restricted to
a finite number of variables.
Hence one finds
\begin{align}
\ch V(p) &= (x_1x_2\cdots)^{p/2} \frac{E_{(0,p)}(x)}{\prod_i (1-x_i)\prod_{i<j}(1-x_ix_j)} \nn\\
& = (x_1x_2\cdots)^{p/2} 
\frac{ \sum_{\eta\in{\cal E}} (-1)^{(|\eta|+r)/2} s_{\eta_{(0,p)}}(x) }{\prod_i (1-x_i)\prod_{i<j}(1-x_ix_j)}. 
\end{align}

\setcounter{equation}{0}
\section{Summary} \label{sec:summary}

The explicit construction of the parafermion and paraboson Fock representations 
was an unsolved problem for many years. 
A solution was in principal offered by means of Green's ansatz~\cite{Green}. 
In practice however, performing Green's ansatz (i.e.\ extracting an irreducible component
in a $p$-fold tensor product) turned out to be difficult, and it has not lead to an explicit general solution. 
The case of finite degrees of freedom, namely the explicit construction of the Fock space 
for $n$ pairs of parafermions and the Fock space for $n$ pairs of parabosons was solved recently~\cite{paraboson,parafermion}. 
For this purpose we used the known relation between the Lie algebra $\so(2n+1)$ and the $n$-parafermion algebra,
between the Lie superalgebra $\osp(1|2n)$ and the $n$-paraboson algebra,
and the relation between parafermion (resp.\ paraboson) Fock spaces and certain unitary
irreducible representations of these algebras.
The main tools used in~\cite{paraboson,parafermion} are:
the decomposition of an induced $\so(2n+1)$ and $\osp(1|2n)$ representation with respect to the compact subalgebra
$\u(n)$, the use of $\u(n)$ GZ-patterns to label basis vectors of the representations, 
the method of reduced matrix elements for $\u(n)$ tensor operators, explicit $\u(n)$ CGCs,
and computational techniques (using a computer algebra package). 
From the point of view of quantum field theory the interesting cases are $n=+\infty$. 
In this paper we have extended the results of~\cite{paraboson,parafermion}
and constructed representations of order~$p$ for an infinite set of parafermions and parabosons.
The corresponding algebras are the infinite rank Lie algebra $\so(\infty)$ and the 
infinite rank Lie superalgebra $\osp(1|\infty)$. The construction of the Fock spaces 
was possible because of the introduction of stable infinite GZ-patterns, the determination
of irreducible $\so(\infty)$ and $\osp(1|\infty)$ actions on them (employing the
corresponding action in the finite rank case) and stability properties of the reduced matrix elements
of the finite $n$ case when $n$ goes to infinity.

Thus in the series of papers~\cite{paraboson,parafermion} and the present one we have given 
solutions to a problem that had been open for a long time. 
An interesting next step would be to investigate the representations of the 
parastatistics algebra~\cite{Loday} consisting of a combined system of $m$
pairs of parafermions and $n$ pairs of parabosons, known to be
related to representations of the orthosymplectic Lie superalgebra $\osp(2m+1|2n)$~\cite{Palev1982}.

\section*{Acknowledgments}
The authors would like to thank Professor T.D.~Palev and Dr.\ S.\ Lievens for their interest.
N.I.~Stoilova was supported by a project from the Fund for Scientific Research -- Flanders (Belgium)
and by project P6/02 of the Interuniversity Attraction Poles Programme (Belgian State -- 
Belgian Science Policy).

\end{document}